\documentclass[%
 reprint,
 superscriptaddress,
 numerical,
 showpacs,
 amsmath,amssymb,
 aps,
 longbibliography,
 prd,
 floatfix
]{revtex4-1}

\usepackage{graphicx}

\usepackage[%
  colorlinks=true,
  urlcolor=blue,
  linkcolor=blue,
  citecolor=blue
]{hyperref}

\newcommand{\ket}[1]{| #1 \rangle}
\newcommand{\bra}[1]{\langle#1 |}

\DeclareMathOperator{\tr}{tr}

\DeclareMathOperator{\Imag}{Im}

\begin{document}

\title{Symmetry classes, many-body zero modes, and supersymmetry in the complex Sachdev-Ye-Kitaev model}

\author{Jan Behrends}
\affiliation{T.C.M. Group, Cavendish Laboratory, University of Cambridge, J.J. Thomson Avenue, Cambridge, CB3 0HE, United Kingdom}
\author{Benjamin B\'{e}ri}
\affiliation{T.C.M. Group, Cavendish Laboratory, University of Cambridge, J.J. Thomson Avenue, Cambridge, CB3 0HE, United Kingdom}
\affiliation{DAMTP, University of Cambridge, Wilberforce Road, Cambridge, CB3 0WA, United Kingdom}

\begin{abstract}
The complex Sachdev-Ye-Kitaev (cSYK) model is a charge-conserving model of randomly interacting fermions.
The interaction term can be chosen such that the model exhibits chiral symmetry.
Then, depending on the charge sector and the number of interacting fermions, level spacing statistics suggests a fourfold categorization of the model into the three Wigner-Dyson symmetry classes.
In this work, inspired by previous findings for the Majorana Sachdev-Ye-Kitaev model, we embed the symmetry classes of the cSYK model in the Altland-Zirnbauer framework and identify  consequences of chiral symmetry originating from correlations across different charge sectors.
In particular, we show that for an odd number of fermions, the model hosts exact many-body zero modes that can be combined into a generalized fermion that does not affect the system's energy.
This fermion directly leads to quantum-mechanical supersymmetry that, unlike explicitly supersymmetric cSYK constructions, does not require fine-tuned couplings, but only chiral symmetry.
Signatures of the generalized fermion, and thus supersymmetry, include the long-time plateau in time-dependent correlation functions of fermion-parity-odd observables:
The plateau may take nonzero value only for certain combinations of the fermion structure of the observable and the system's symmetry class. 
We illustrate our findings through exact diagonalization simulations of the system's dynamics. 
\end{abstract}

\maketitle

\section{Introduction}

The holographic principle is a conjectured relationship between gravitational and lower-dimensional conformal field theories~\cite{Susskind:1995hz,Gubser:1998id,Witten:1998ko,Maldacena:1999js}.
It has been an intriguing approach to explore non-Fermi liquid phases in strongly correlated quantum systems that are characterized by the absence of quasiparticles~\cite{Hartnoll:2012ce,Sachdev:2012jw}.
The Sachdev-Ye-Kitaev (SYK) model that describes randomly interacting degrees of freedom is a toy model for such a non-Fermi liquid phase~\cite{Sachdev:1993hv,Kitaev2015}, often referred to as a strange metal.
Originally formulated by Sachdev and Ye in terms of randomly interacting $\mathrm{SU}(M)$ spins~\cite{Sachdev:1993hv}, the model has seen a recent surge of interest due to the insight by Kitaev linking a Majorana fermion variant of the model to a gravitational dual~\cite{Kitaev2015}:
a two-dimensional nearly anti-de Sitter space, which arises for near-extremal black holes~\cite{NavarroSala:2000id,Sachdev:2010kq,Kitaev2015}.

Similarly to black holes, the SYK model is maximally chaotic~\cite{Kitaev2015}.
This means that any inserted information is distributed over all degrees of freedom with maximal efficiency, akin to the maximal scrambling of information in black holes~\cite{Shenker:2014ct,Maldacena:2016gp}.
The scrambling of quantum information is reflected in the exponential decay of out-of-time-order correlation functions~\cite{Kitaev2015,Maldacena:2016gp} whose study sparked renewed interest in quantum chaos~\cite{Roberts:2016by,Rozenbaum:2017br,Patel:2017fq,Bagrets2017,Altland2018,Micklitz:2019kt}.

The variant of the model with randomly interacting complex fermions shares many properties with the Majorana version, including a nonzero entropy density at zero temperature, but it additionally conserves a $\mathrm{U}(1)$ charge~\cite{Sachdev:2015dp,Gu2019}.
The model exhibits compressible states, i.e., the charge density can be tuned by a conjugate chemical potential~\cite{Sachdev:2015dp}.
It is further characterized by a spectral asymmetry between the particle and hole excitations, which is related to the charge and entropy densities~\cite{Parcollet:1998bd,Georges:2011jz}.
The candidate holographic dual of the charge-conserving complex-fermion SYK model is a charged black hole~\cite{Sachdev:2015dp}.
In this dual, the spectral asymmetry is replaced by the strength of the electric field on the horizon and a similar relation between charge and entropy densities can be found~\cite{Sen:2005ef,Sen:2008en,Faulkner:2011da}.
In the following, we refer to Majorana fermion model as the SYK model, and the complex fermion model as the cSYK model.

The above properties of the cSYK and SYK models are understood in the thermodynamic and low-energy limit.
Numerical studies provide complementary insights to the problem through the access to the entire spectrum and all eigenfunctions.
However, they are only feasible up to a small number of interacting degrees of freedom and therefore miss some of the low-energy features relevant in the thermodynamic limit~\cite{Fu:2016kz,Hosur:2016bt}.
Mesoscopic (c)SYK variants with a moderate number of fermions may bridge the gap between these limits.
Several condensed-matter systems have been suggested to have low-energy physics effectively described by randomly interacting Majorana~\cite{Pikulin:2017js,Chew:2017fn} or complex fermions~\cite{Chen:2018ho,Altland:2019ck}, thereby providing platforms for analog (c)SYK quantum simulations.
Conversely, digital, qubit-based, quantum simulation has also been proposed for capturing the physics of the model via the measurement of time-dependent correlation functions~\cite{Alvarez2017,Luo:2019fp,Babbush:2019eh}.

In the mesoscopic variants of the cSYK and SYK model, it is possible to access observables complementary to thermodynamic quantities, such as the level spacing statistics.
As pointed out by You \textit{et al.}, the level spacing statistics in the SYK model are solely determined by the number of Majorana fermions and follow the three Wigner-Dyson symmetry classes with an eightfold periodicity in the Majorana fermion number~\cite{You:2017jj}.
This eightfold pattern was later understood to originate from a symmetry classification beyond Wigner-Dyson, namely the realization of the eight real Altland-Zirnbauer classes~\cite{Altland:1997cg} in the SYK model~\cite{Behrends:2019jc}. 
This has several structural consequences, including many-body zero modes~\cite{Behrends:2019jc} and quantum mechanical supersymmetry~\cite{Behrends2019}, and signatures
beyond level spacing statistics, in particular in fermion-parity-odd correlation functions~\cite{Behrends:2019jc,Cotler:2017fx,Behrends2019}.

For the cSYK model with chiral symmetry, You \textit{et al.\@} found that the number $N$ of complex fermions and the charge quantum number $q$ together determine a fourfold pattern for the three Wigner-Dyson level spacing statistics. 
This leads us to the main questions of this work:
Is there a symmetry classification beyond Wigner-Dyson  behind this fourfold pattern and, if so, what are its structural consequences and signatures?

We show that the symmetry classes relevant for the chiral-symmetric cSYK model are the four real Altland-Zirnbauer classes with only one antiunitary symmetry, namely classes AI, AII, C and D. (The complementary four, not appearing in this model, are characterized by the presence of two antiunitaries.) 
A key structural consequence we find is that, for odd $N$, the system supports many-body zero modes: 
fermion-parity-odd objects that exactly commute with the Hamiltonian
~\footnote{Here and in what follows, we reserve the term ``many-body zero modes'' to such fermion-parity-odd objects, but we note that analogous, parity-even, objects are also supported by the cSYK model.}.

We also show that the the many-body zero modes of the odd-$N$, chiral-symmetric, cSYK model can be combined into a generalized fermion, and that the presence of this means that the system is supersymmetric.
We use supersymmetry (SUSY) in the sense of supersymmetric quantum mechanics, which is characterized by the number and structure of Hermitian supercharges that anticommute and square to the Hamiltonian~\cite{Nicolai:1976gf,Witten:1982cs,Junker:1996ej,Fendley:2003ch,Fendley:2003ef}.
While the number of supercharges is always two (unlike in the Majorana SYK model~\cite{Behrends2019}), the structure of the supercharges depends on the symmetry class at hand.

This directly translates into signatures in correlation functions. 
In particular, correlators of fermion-parity-odd objects, including the single-particle Green's function, may exhibit a plateau at long times.
Whether this plateau is nonzero, depends on the system's symmetry class (class C or D for odd $N$) and the fermion structure of the object under consideration.

We emphasize that the symmetry classification, the structural features, and the signatures we find all crucially rely on the presence of chiral symmetry. 
Given that cSYK models can also be defined without this symmetry, we shall also discuss different ways to generate couplings resulting in chiral-symmetric cSYK Hamiltonians.

This work is organized as follows:
In Sec.~\ref{sec:model}, we discuss the symmetry classification of the chiral-symmetric cSYK model and summarize known features of the level spacing statistics.
In Sec.~\ref{sec:zero_modes}, we identify the model's many-body zero modes, including the aforementioned generalized fermion.
In Sec.~~\ref{sec:susy}, we use this generalized fermion to show that the cSYK model is supersymmetric and discuss the structure of the supercharges.
The zero modes and SUSY have direct consequences for various correlation functions, as we demonstrate explicitly in Sec.~\ref{sec:signatures}, and illustrate by numerical results for time-dependent correlation functions.
We conclude in Sec.~\ref{sec:conclusion} and give an outlook on possible extensions of our results.

\section{Model and symmetry classification}
\label{sec:model}

Throughout this work, we consider the cSYK Hamiltonian~\cite{Sachdev:1993hv,Sachdev:2015dp}
\begin{equation}
 H = \sum\limits_{ijkl=1}^N J_{ij,kl} c_i^\dagger c_j^\dagger c_k c_l - \mu \sum\limits_{i=1}^N \left(c_i^\dagger c_i - \frac{1}{2} \right)
 \label{eq:hamiltonian}
\end{equation}
of $N$ complex spinless fermions that interact via a four-body term with structureless complex couplings $J_{ij,kl}$.
For convenience, we include a chemical potential $\mu$.
The fermions $c_j$ obey the usual relations $\{ c_i, c_j \} = 0$ and $\{ c_i^\dagger , c_j \} = \delta_{ij}$.
To ensure that $H$ is a Hermitian operator, the couplings need to satisfy $J_{ij,kl} = J_{lk,ji}^*$.
Due to fermionic anticommutation, only the antisymmetric part of the couplings contributes, thus we can safely set
\begin{align}
 J_{ij,kl} = - J_{ji,kl} = - J_{ij,lk} = J_{ij,kl} .
\end{align}
This model conserves charge
\begin{equation}\label{eq:chargeoperator}
 Q = \sum_{i=1}^N \left( c_i^\dagger c_i - \frac{1}{2} \right)
\end{equation}
with eigenvalues $q_n=n-N/2$, and $n=0,1,\ldots ,N$.
The charge operator $Q$ thus has integer eigenvalues for even $N$, and half-integer eigenvalues for odd $N$.
Since $[ H, Q] =0$, the eigenstates of $H$ can be labeled by the charge $q$,
\begin{align}
 H \ket{\psi_\mu^q} = \varepsilon_\mu^q \ket{\psi_\mu^q} , & & Q \ket{\psi_\mu^q} = q \ket{\psi_\mu^q}
\end{align}
and the Hamiltonian can be brought into a block-diagonal form.
We refer to each of the blocks as a charge sector, each characterized by their charge $q$.

Extra structure in the couplings may imply additional symmetries.
The key extra structure relevant to us arises if the interaction terms $c_i^\dagger c_j^\dagger c_k c_l$ appear in antisymmetrized combinations of the creation and annihilation operators. 
This renders the Hamiltonian invariant under $c_i^\dagger \leftrightarrow c_i$ and $J_{ij,kl} \leftrightarrow J_{ij,kl}^*$~\cite{You:2017jj,Gu2019}, i.e., we have $[H,\mathcal{S}]=0$ with the \emph{antiunitary} chiral symmetry~\cite{You:2017jj,Ludwig:2016cd} $\mathcal{S}$ effecting
\begin{align}
 \mathcal{S} c_j \mathcal{S}^{-1} = c_j^\dagger , & &
 \mathcal{S} c_j^\dagger \mathcal{S}^{-1} = c_j .
 \label{eq:chiral_symmetry}
\end{align}
In Eq.~\eqref{eq:chargeoperator}, we defined the charge operator such that chiral symmetry anticommutes with charge $ \{ \mathcal{S}, Q \} = 0$.

Before turning to analyzing the consequences of chiral symmetry, we comment on a few ways in which the required coupling structure may arise. 
Starting from the original Hamiltonian Eq.~\eqref{eq:hamiltonian}, upon applying $\mathcal{S}$, we find
\begin{align}
 \mathcal{S} H \mathcal{S}^{-1} &= H + 2 \mu Q - 2 \sum_{ijk} J_{ij,ki} \left( c_j^\dagger c_k - c_k c_j^\dagger \right).
 \label{eq:hamiltonian_chiral_trafo}
\end{align}
The Hamiltonian respects chiral symmetry if the last two terms sum to zero. 
If the couplings satisfy for example
\begin{equation}
 \sum_i J_{ij,ki} = \frac{\varepsilon_0}{2} \delta_{jk} ,
\end{equation}
then Eq.~\eqref{eq:hamiltonian_chiral_trafo} simplifies to
\begin{align}
 \mathcal{S} H \mathcal{S}^{-1} &= H + 2 (\mu - \varepsilon_0) Q .
 \label{eq:chiral_hamiltonian}
\end{align}
In this case, an appropriate chemical potential $\mu = \varepsilon_0$ fixes chiral symmetry to be exact.
Alternatively, one could use a more relaxed definition of chiral symmetry with $\mu \neq \varepsilon_0$.
Then, the Hamiltonian does not transform to itself under chiral symmetry, but it is shifted by an energy that depends solely on the charge sector.
In this work, we focus on the presence of exact chiral symmetry, i.e., $\mathcal{S} H \mathcal{S}^{-1} = H$, although some of the features we discuss are also present using the more relaxed definition.

One notable example of couplings satisfying chiral symmetry are those generated by a supersymmetric construction of the cSYK model using the nilpotent supercharges~\cite{Fu:2017hl}
\begin{equation}\label{eq:explSUSY}
 \mathcal{Q}_s = \sum_{ijk} C_{ijk} c_i c_j c_k
\end{equation}
with antisymmetric coefficients $C_{ijk}$ and $\mathcal{Q}_s^2 =0$.
Under chiral symmetry, $\mathcal{Q}_s$ transforms as~\cite{Fu:2017hl,Li:2017bi,Kanazawa:2017be}
\begin{align}
 \mathcal{S} \mathcal{Q}_s \mathcal{S}^{-1} = - \mathcal{Q}_s^\dagger, & &
 \mathcal{S} \mathcal{Q}_s^\dagger \mathcal{S}^{-1} = - \mathcal{Q}_s, & &
\end{align}
such that the Hamiltonian $H = \lbrace  \mathcal{Q}_s , \mathcal{Q}_s^\dagger \rbrace$ respects chiral symmetry.
In Sec.~\ref{sec:susy}, we describe, in some sense, the reverse of the above statement: 
In the presence of chiral symmetry, the cSYK Hamiltonian~\eqref{eq:hamiltonian} with odd $N$ can always be written in terms of two supercharges, albeit those have more intricate structure than Eq.~\eqref{eq:explSUSY}.

We now turn to the analysis of the consequences of chiral symmetry. 
The operator $\mathcal{S}$ can be explicitly constructed by employing a Jordan-Wigner transformation~\cite{Jordan:1928kq,You:2017jj}.
Writing the creation and annihilation operator as spin-$1/2$ raising and lowering  operators $\sigma_j^\pm = (\sigma^x_j \pm i \sigma^y_j)/2$ attached to a string of spin operators,
\begin{align}
 c_i = \prod_{j<i} \sigma^z_j \sigma^-_i , & &  c_i^\dagger = \prod_{j<i} \sigma^z_j \sigma^+_i ,
 \label{eq:jordan_wigner}
\end{align}
the creation and annihilation operator are both represented by purely real matrices.
Following Ref.~\onlinecite{You:2017jj}, we split the operator $\mathcal{S} = \mathcal{U}\mathcal{K}$ into a unitary part $\mathcal{U}$ and complex conjugation $\mathcal{K}$.
Since the representation of $c_i$ in terms of spin operators is real, $c_i^\dagger = \mathcal{S} c_i \mathcal{S}^{-1} = \mathcal{U} c_i \mathcal{U}^\dagger$ in this basis.
This transformation can be achieved by the unitary matrix consisting of an alternating string of $\sigma^x_j$ and $\sigma^y_{j+1}$~\cite{Fidkowski:2010ko,Fidkowski:2011dh,You:2017jj}
\begin{equation}
 \mathcal{U} = \begin{cases}
 \sigma^x_1 \sigma^y_2 \cdots \sigma^x_{N-1}\sigma^y_N & \text{ for even }N \\
 \sigma^x_1 \sigma^y_2 \cdots \sigma^y_{N-1}\sigma^x_N & \text{ for odd }N .
\end{cases}
\end{equation}
The square of $\mathcal{S}$ can be easily deduced from the number of $\sigma^y_j$ operators in $\mathcal{U}$, namely
\begin{equation}
 \mathcal{S}^2 = \mathcal{U} \mathcal{U}^* =
 \begin{cases} +1 & \text{ for } N \mod 4 = 0,1 \\ -1 &  \text{ for } N \mod 4 = 2,3 . \end{cases}
\end{equation}
For $N$ even, $\mathcal{S}$ conserves the fermion parity $P = (-1)^{Q+N/2}$, $[ \mathcal{S},P ] =0$, whereas it changes the fermion parity for odd $N$, $\{ \mathcal{S},P \} = 0$~\cite{Behrends:2019jc}.
This allows us to discuss four distinct cases depending on $N\mod 4$.
We either have $N$ even (and hence $[\mathcal{S},P]=0$) with $\mathcal{S}^2 = \pm1$, or $N$ odd (and hence $\{ \mathcal{S},P \} = 0$) with $\mathcal{S}^2 = \pm 1$.

\begin{figure}
\includegraphics{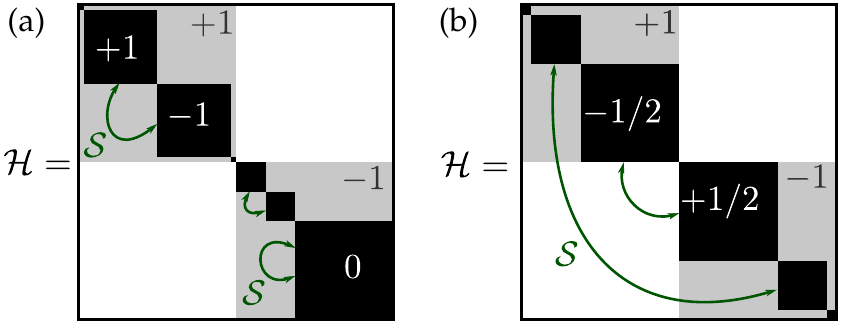}
\caption{Schematic block-structure of the cSYK Hamiltonian.
The Hamiltonian consists of two parity blocks with parity eigenvalue $p = \pm 1$ (gray blocks with $p$ shown in the upper right corners).
Each of the parity blocks consists of several subblocks for the different charge sectors (black blocks with charge eigenvalues $q$ illustrated by the white numbers).
Only matrix elements in the black blocks are nonzero. The action of the chiral symmetry $\mathcal{S}$ is illustrated by the arrows. 
(a) When $N$ is even, chiral symmetry $\mathcal{S}$ commutes with fermion parity and acts within the parity sectors. It does, however, anticommute with the charge operator $Q$ and thus exchanges sectors $q$ and $-q$. Only $q=0$ is left invariant.
(b) When $N$ is odd, chiral symmetry exchanges parity sectors.
It exchanges charge sectors $q$ and $-q$ and leaves no block invariant.
}
\label{fig:chiral_schematics}
\end{figure}

Below, we first briefly describe the how the $\mathcal{S}^2 = \pm 1$ cases for $N$ even fit into the SYK symmetry classification of Ref.~\onlinecite{Behrends:2019jc}, highlighting also the qualifications required due to charge conservation. 
Our considerations here lead to the same two classes as the ones identified in Ref.~\onlinecite{You:2017jj} via level spacing statistics.  
Then, we turn to the main focus of this work:
the classes arising for odd $N$.
Here, while level spacing statistics alone does not distinguish between $\mathcal{S}^2 = \pm 1$, we find that two distinct classes arise in the two cases. 
The identification of the symmetry classes is aided by considering the block structure of the Hamiltonian and the interrelation of the different blocks under the action of $\mathcal{S}$, as illustrated in in Fig.~\ref{fig:chiral_schematics}. 
We organize the Hamiltonian into two parity blocks, and within those, different charge sectors. 

When $N$ is even, the $-q$ and $q$ charge sectors have the same fermion-parity eigenvalue $p =(q+N/2) \mod 2$. 
For $q\neq0$ they are exchanged under $\mathcal{S}$. 
Consistency with the SYK classification of Ref.~\onlinecite{Behrends:2019jc} leads us to associate to $\mathcal{S}^2=+1$ and $\mathcal{S}^2=-1$ the Cartan labels AI (orthogonal class) and AII (symplectic class), respectively: 
As in our cSYK model, in the Majorana SYK case the antiunitary symmetry present for $(2N)\mod 8 = 0,4$ Majoranas (i.e., $N\mod 4 = 0,2$ complex fermions) conserves fermion parity, and in this sense is time-reversal-like. 
Since this is the only antiunitary symmetry in this case, the orthogonal and symplectic classes are the natural association. 
In the cSYK model, however, each parity sector consists of several subblocks with different charge eigenvalues $q$; we thus have, in each parity block, class AI (AII) with charge conservation supplying an extra unitary symmetry. 
This masks the level repulsion characteristics of class AI (AII) in all the charge sectors but $q=0$, which is left invariant by $\mathcal{S}$.
Specifically, as explained by You \textit{et al.}, the level spacing statistics in all $q\neq 0$ charge sectors follow the Gaussian unitary ensemble~\cite{You:2017jj}; 
however, the level spacing statistics in the $q=0$ subblock follow the Gaussian orthogonal ensemble when $\mathcal{S}^2=+1$ and Gaussian symplectic ensemble when $\mathcal{S}^2 =-1$, matching classes AI and AII, respectively. 

For odd $N$, the charge sectors $-q$ and $q$ have opposite fermion parity.
Thus, chiral symmetry acts across opposite parity sectors. 
In particular, it leaves none of the charge sectors invariant, hence there is no charge sector in which it could influence level spacing statistics;
these follow, for all $q$, the Gaussian unitary ensemble~\cite{You:2017jj,Garcia:2016il}.
Due to these facts, the resulting symmetry classes are D ($\mathcal{S}^2 = +1$) and C ($\mathcal{S}^2=-1$), where the specific association is done consistently with the symmetry classification of the Majorana SYK model~\cite{Behrends:2019jc}.
We summarize the symmetry classes and level spacing statistics in Table~\ref{tab:symmetry_classes}.

The consequences of chiral symmetry connecting different charge and parity sectors remain largely unexplored. 
It is these consequences that we focus on in the following sections.
We shall focus on odd $N$, where consequences will be shown to include the presence of many-body zero modes and quantum-mechanical SUSY, as well as signatures in fermion-parity-odd dynamical correlation functions. 
When $N$ is even, certain parity-even correlation functions away from the $q=0$ charge sector may also exhibit signatures of the respective symmetry class due to mechanisms analogous to those we shall describe for the odd-$N$ case. 
We shall briefly comment on these even-$N$ signatures in the outlook at the very end of this work.

\begin{table}\label{tab:classes}
\centering
\begin{tabular}{c|cccc}
 \toprule
  $N \mod 4$ & $0$ & $1$ & $2$ & $3$ \\
  \colrule
  $\mathcal{S} P \mathcal{S}^{-1}$ & $+P$ & $-P$ & $+P$ & $-P$ \\
  $\mathcal{S}^{2}$ & $+1$ & $+1$ & $-1$ & $-1$ \\
  \colrule
 Cartan  & AI & D & AII & C \\
 \colrule
 $q = 0$ & GOE & & GSE & \\
 $q \neq 0$ & GUE & GUE & GUE & GUE \\
 \botrule
\end{tabular}
\caption{Altland-Zirnbauer symmetry classes of the cSYK model.
The sign $s=\pm 1$ in  $\mathcal{S} P \mathcal{S}^{-1} = s P$ together with the value of $\mathcal{S}^2 = \pm 1$ gives four different symmetry classes for the four values of $N\mod 4$.
Only the $q=0$ sector is left invariant under chiral symmetry, such that it is the only charge sector whose level spacing statistics follow the corresponding Gaussian orthogonal (GOE) and symplectic ensemble (GSE), whereas the level spacing statistics in all other charge sectors follow the Gaussian unitary ensemble (GUE).
}
\label{tab:symmetry_classes}
\end{table}

\section{Zero modes}
\label{sec:zero_modes}

In this section, we exploit the interrelation between the opposite charge sectors $q$ and $-q$ to construct many-body zero modes, operators that commute with the Hamiltonian but anticommute with the fermion parity.
A key ingredient for this will be seen to be the anticommutation between $\mathcal{S}$ and fermion parity, thus such zero modes are present only for odd $N$. 

Using $H \mathcal{S} = \mathcal{S} H$, we realize that the states $\ket{\psi_\mu^q}$ and $\ket{\mathcal{S} \psi_\mu^{q}}$ are both eigenstates of the Hamiltonian with the same energy $\varepsilon_\mu^q = \varepsilon_\mu^{-q}$,
\begin{align}
  H \ket{\mathcal{S} \psi_\mu^q} &= \mathcal{S} H \ket{\psi_\mu^q} = \varepsilon_\mu^q  \ket{\mathcal{S} \psi_\mu^q} ,
\end{align}
but opposite charge eigenvalues $q$ and $-q$,
\begin{align}
  Q \ket{\mathcal{S} \psi_\mu^q} &= - \mathcal{S} Q \ket{\psi_\mu^q} = -q  \ket{\mathcal{S} \psi_\mu^q} .
\end{align}
We adopt the following convention between eigenstates in opposite charge sectors: $\ket{\psi_\mu^{-q}} \equiv \ket{\mathcal{S} \psi_\mu^{q}}$ (with $q>0$). 

In the spirit of Ref.~\onlinecite{Behrends:2019jc}, we construct eigenoperators of the Hamiltonian ($q > 0$)
\begin{align}\label{eq:eigenops}
 \mathcal{O}_{\mu\nu}^q &= \ket{\psi_\mu^q}\bra{\psi_\nu^{-q}} ,\\
 [ H, \mathcal{O}_{\mu\nu}^q ] &= (\varepsilon_\mu^q - \varepsilon_\nu^{-q}) \mathcal{O}_{\mu\nu}^q
\end{align}
that change the charge by $2q$, which for odd $N$ takes values $2q=1,3,\ldots,N$. 
Since the energies in opposite charge sectors are degenerate $\varepsilon_\mu^q = \varepsilon_\mu^{-q}$, the eigenoperator $\mathcal{O}_{\mu\mu}^q$ commutes with the Hamiltonian, while due to $2q$ being odd, it anticommutes with fermion parity:
It is an odd-parity many-body zero mode.

Using these zero modes as building blocks, we next define a generalized (many-body) fermion:
\begin{equation}
 d^\dagger = \sum_{\mu,q>0} \ket{\psi_\mu^q} \bra{\psi_\mu^{-q}}.
\end{equation}
This satisfies $\{ d ,d^\dagger \} = 1$, $d^2=0$ and has the property that $d$ decreases the charge of any state in a charge sector $q>0$ by $2q$, whereas $d^\dagger$ increases the charge of a state in a charge sector $-q<0$ by $2q$. 
Both operators $d$ and $d^\dagger$ commute with the Hamiltonian and anticommute with fermion parity since their constituent eigenoperators $\mathcal{O}_{\mu\mu}^q$ all do so individually.

Under the application of chiral symmetry, creation and annihilation operators of the generalized fermion transform into another via
\begin{align}
 \mathcal{S} d^\dagger \mathcal{S}^{-1} = \pm d & &
 \mathcal{S} d \mathcal{S}^{-1} = \pm d^\dagger ,
 \label{eq:chiral_fermions}
\end{align}
where the sign is that of $\mathcal{S}^2 = \pm 1$.
When $\mathcal{S}^2 = -1$, this transformation is different from the action of chiral symmetry on the complex fermions $c_j$, Eq.~\eqref{eq:chiral_symmetry}.
This implies that the generalized fermion $d$ cannot have any weight in the complex-fermion annihilation operator $c_j$ in its expansion in operator space.

To investigate the decomposition of $d$ into the constituent fermions $c_j$, we first expand it in terms of products of Majorana operators.
Each complex fermion can be split up into two Majorana operators $\gamma_j$.
Here, we use $\gamma_{2j} = c_j^\dagger + c_j$ and $\gamma_{2j+1} = i (c_j - c_j^\dagger)$~\cite{Kitaev:2001gb}.
These $2N$ Majorana operators are Hermitian, obey the anticommutation relation $\{ \gamma_i,\gamma_j \} = 2\delta_{ij}$, and commute with chiral symmetry, $[\mathcal{S},\gamma_j] = 0$.
We employ the Majorana operators to construct a basis that spans all operators that act on the Hilbert space of $N$ complex fermions.
In particular, the products of $n_a$ Majorana operators~\cite{Goldstein:2012ci}
\begin{equation}
 \Upsilon_a = i^{n_a (n_a-1)/2} \gamma_{i_1 (a)} \gamma_{i_2 (a)} \cdots \gamma_{i_{n_a} (a)}
 \label{eq:upsilon}
\end{equation}
are Hermitian, unitary, and orthonormal with respect to the trace, $\tr [ \Upsilon_a \Upsilon_b ] = 2^N \delta_{ab}$.
Any operator acting on the same Hilbert space can be expanded in terms of the $\Upsilon_a$ with complex coefficients.
Using $\mathcal{S} \gamma_j \mathcal{S}^{-1} = \gamma_j$, we find~\cite{Behrends2019}
\begin{equation}
 \mathcal{S} \Upsilon_a \mathcal{S}^{-1}  = (-1)^{n_a (n_a-1)/2} \Upsilon_a .
 \label{eq:upsilon_trafo}
\end{equation}
The $\Upsilon_a$ do not conserve charge, but may conserve fermion parity $P$.
In particular, $\Upsilon_a$ with even (odd) $n_a$ conserve (change) fermion parity, 
\begin{equation}
 P \Upsilon_a P^{-1} = (-1)^{n_a} \Upsilon_a .
\end{equation}
As $d^{(\dagger)}$ changes the fermion parity,
only $\Upsilon_a$ with odd $n_a$ can contribute to $d^{(\dagger)}$.
Similarly, only those $\Upsilon_a$ that transform with the same sign under $\mathcal{S}$ as $d^{(\dagger)}$ can have nonzero weight in the expansion of $d^{(\dagger)}$.
This implies the only contribution to the sum
\begin{align}
 d = \sum_a v_a \Upsilon_a , & &  d^\dagger = \sum_a v_a^* \Upsilon_a
\end{align}
are terms with $n_a = 4n+1$ when $\mathcal{S}^2 = +1$ (class D) and terms with $n_a = 4n+3$ when $\mathcal{S}^2 =-1$ (class C), with $n$ a nonnegative integer.
This is analogous to the pattern found for classes D and C for the Majorana SYK model, as discussed in Ref.~\onlinecite{Behrends2019}.

To understand the consequences of the expansion in terms of products of Majorana fermions better, we take advantage of charge conservation.
We first note that linear combinations $\bar{\Upsilon}_a = \sum_b\Upsilon_b U_{ba}^*$ with the unitary matrix $U$ are also orthonormal with respect to the trace,
\begin{align}
 ( \bar{\Upsilon}_a , \bar{\Upsilon}_b )
 &= \frac{1}{2^N} \tr [ \bar{\Upsilon}_a^\dagger \bar{\Upsilon}_b ] \nonumber \\
 &= \frac{1}{2^N} \sum_{a'b'}U_{a'a} U_{b'b}^* \tr [ \Upsilon_{a'} \Upsilon_{b'} ] =\delta_{ab} ,
\end{align}
where we denoted the trace inner product of two matrices by $(A,B)$~\footnote{Unlike the operators $\Upsilon_a$, the $\bar{\Upsilon}_a$ are generally non-Hermitian, such that the Hermitian conjugate of $A$ is necessary in the definition of the trace inner product $(A,B)= \tr[A^\dagger B]/\dim A$.}.
We only consider matrices $U$ with $U_{ab} =0$ when $n_a\neq n_b$, such that each $\bar{\Upsilon}_a$ contains a well-defined number of Majorana fermions.
The operators $\bar{\Upsilon}_a$ transform under chiral symmetry as
\begin{equation}
 \mathcal{S} \bar{\Upsilon}_a \mathcal{S}^{-1} = (-1)^{n_a(n_a-1)/2} \bar{\Upsilon}_a^\dagger .
\end{equation}
To take advantage of charge conservation, we aim to construct $\bar{\Upsilon}_a$ and $\bar{\Upsilon}_a^\dagger$ that are eigenoperators of the charge operator,
\begin{equation}
 [ Q , \bar{\Upsilon}_a^\dagger ] = q_a \bar{\Upsilon}_a^\dagger,
 \label{eq:commutation}
\end{equation}
such that $\bar{\Upsilon}_a^\dagger$ increases the charge of a given state by $q_a$.
After multiplying Eq.~\eqref{eq:commutation} from the left by $\Upsilon_b$ and taking the trace, we obtain
\begin{align}
 \frac{1}{2^N} \sum_{a'} \tr [ \Upsilon_b [ Q , \Upsilon_{a'} ] ] U_{a'a}
 &=  \frac{q_a}{2^N} \sum_{a'} \tr[ \Upsilon_b \Upsilon_{a'} ] U_{a'a} \nonumber \\
 &= q_a U_{ba},
\end{align}
where we used the orthogonality of the $\Upsilon_a$ in the last step.
The $a$th column of $U$ is thus the eigenvector with eigenvalue $q_a$ of the Hermitian and skew-symmetric matrix $\mathcal{C}$ with elements
\begin{equation}
 \mathcal{C}_{ab} = \frac{1}{2^N} \tr[ \Upsilon_a [ Q , \Upsilon_{b} ] ] = - \mathcal{C}_{ba} .
\end{equation}
This approach to obtain eigenoperators of $Q$ is similar to the procedure described in Ref.~\onlinecite{Goldstein:2012ci} for obtaining eigenoperators of interacting Majorana Hamiltonians.
Each set of $\binom{2N}{n_a}$ different $\bar{\Upsilon}_a$ with fixed $n_a$ consists of  different subsets of
\begin{equation}
 N_{n_a q_a} \equiv \binom{N}{\frac{n_a + q_a}{2}} \binom{N}{\frac{n_a - q_a}{2}}
\end{equation}
operators that change the charge by $q_a$.
The sum over all $q_a$ gives back the $\binom{2N}{n_a}$ operators, which can be verified using Vandermonde's identity.

For small $n_a$, we can find transparent constructions of these operators, e.g., for $n_a \le 2$,
\begin{subequations}\begin{align}
 \bar{\Upsilon}_a|n_a=0 :& ~1 \\
 \bar{\Upsilon}_a|n_a=1 :& ~\{ \sqrt{2} c_i \} , \{ \sqrt{2} c_i^\dagger \} \\
 \bar{\Upsilon}_a|n_a=2 :& ~\{ 2 c_i c_j | i < j \},\{ 2 c_i^\dagger c_j^\dagger | i < j \} ,\{ 2 c_i^\dagger c_j | i \neq j \},\nonumber \\
 &~ \{ c_i^\dagger c_i - c_i c_i^\dagger \} .
\end{align}\end{subequations}
As operators with $n_a = 1$ do not have any weight on the zero mode $d$ in class C, the expansion of $d$ does not contain single creation and annihilation operators.
The only terms contributing to $d$ that decrease the charge by one are of the form $c_i^\dagger c_j c_k$ (and higher order in $c_j$).
Similarly, operators with $n_a = 3$ do not have any weight on the zero mode in class D, and the expansion does not contain terms $c_i^\dagger c_j c_k$; it does, however, contain single terms $c_i$.
Instead of combing products of creation and annihilation operators, we shall use the operators $\bar{\Upsilon}_a$ and $\bar{\Upsilon}_a^\dagger$ with well-defined $n_a$ and $q_a$ in the following.
Unless stated otherwise, $q_a >0$ so $\bar{\Upsilon}_a^\dagger$ increases the charge by $q_a$ and $\bar{\Upsilon}_a$ decreases the charge by $q_a$.

\section{Supersymmetry}
\label{sec:susy}

The generalized fermion operators $d$ and $d^\dagger$ constructed in the previous section can be promoted to supercharges.
Thus, as we show in the following, when $N$ is odd, chiral symmetry in itself implies quantum-mechanical SUSY for the cSYK model.
This is in contrast to explicitly supersymmetric cSYK constructions~\cite{Fu:2017hl} [see also Eq.~\eqref{eq:explSUSY}], which imply relations between the Hamiltonian couplings $J_{ij,kl}$ beyond chiral symmetry,
so that $N$ need not be odd for SUSY to be present.

Supersymmetric quantum mechanics is characterized by $\mathcal{N}$, the number of Hermitian supercharges that anticommute and square to the Hamiltonian~\cite{Nicolai:1976gf,Witten:1982cs,Wess1992,Junker:1996ej}, and furnish the representation of the supercharges in terms of complex fermions~\cite{Fendley:2003ch,Fendley:2003ef,Kanazawa:2017be}.
After shifting all energies by a constant such that all $\varepsilon_\mu^q \ge 0$, we construct a supercharge via
\begin{equation}
 \mathcal{Q}^\dagger = \sum_{\mu,q>0} \sqrt{\varepsilon_\mu^q} \ket{\psi_\mu^q} \bra{\psi_\mu^{-q}},
\end{equation}
in close analogy to the supercharges in the Majorana SYK model considered in Ref.~\onlinecite{Behrends2019}.
The operators $\mathcal{Q}$ and $\mathcal{Q}^\dagger$ are both nilpotent $\mathcal{Q}^2 = {\mathcal{Q}^\dagger}^2 = 0$, and their anticommutator equals the Hamiltonian
\begin{equation}
 H = \{ \mathcal{Q}^\dagger, \mathcal{Q} \} .
\end{equation}
This implies that the two linearly independent Hermitian supercharges $\bar{\mathcal{Q}}_1 = \mathcal{Q} + \mathcal{Q}^\dagger$ and $\bar{\mathcal{Q}}_2 = i(\mathcal{Q} - \mathcal{Q}^\dagger)$ satisfy $\{ \bar{\mathcal{Q}}_i ,\bar{\mathcal{Q}}_j \} = 2 H \delta_{ij}$.
Thus, we find $\mathcal{N}=2$ supersymmetry.

\begin{table}
\begin{tabular}{c||c|c|c}
 \toprule
 $N \mod 4$ & Cartan &  $\mathcal{S} \mathcal{Q} \mathcal{S}^{-1}$ & fermion structure \\
 \colrule
 $1$ & D &  $+\mathcal{Q}^\dagger$ & $n_a=4n+1$ \\
 $3$ & C & $-\mathcal{Q}^\dagger$ & $n_a=4n+3$ \\
 \botrule
\end{tabular}
\caption{Transformation properties and fermion expansion structure of the supercharges for the two odd-$N$ symmetry classes (classes C and D).
The generalized fermion $d$ has the same transformation and expansion properties.}
\label{tab:transformation}
\end{table}

The supercharges $\mathcal{Q}$ and $\mathcal{Q}^\dagger$ share important features with the generalized fermion operators.
In particular, acting on any given charge sector that they do not annihilate, $d$ and $\mathcal{Q}$ ($d^\dagger$ and $\mathcal{Q}^\dagger$) change the charge the same way. 
Therefore, as do $d$ and $d^\dagger$, both $\mathcal{Q}$ and $\mathcal{Q}^\dagger$ change the fermion parity,
\begin{align}
 \{ \mathcal{Q},P \} = 0 , & &  \{ \mathcal{Q}^\dagger,P \} = 0 .
\end{align}
Chiral symmetry transforms the two operators into each other,
\begin{align}\label{eq:SandQ}
 \mathcal{S} \mathcal{Q}^\dagger \mathcal{S}^{-1} = \pm \mathcal{Q}, & & 
 \mathcal{S} \mathcal{Q} \mathcal{S}^{-1} = \pm \mathcal{Q}^\dagger,
\end{align}
with the sign given by $\mathcal{S}^2 = \pm 1$; cf.\ Eq.~\eqref{eq:chiral_fermions}.
We summarize the key features of the supercharges in Table~\ref{tab:transformation}.

For the same reasons as for the generalized fermion operators (Sec.~\ref{sec:zero_modes}), the supercharges can only contain those $\bar{\Upsilon}_a$ with the same transformation properties under $P$ and $\mathcal{S}$.
This implies that in the expansion
\begin{equation}
 \mathcal{Q}^\dagger = \sum_a u_a \bar{\Upsilon}_{a}^\dagger
\end{equation}
only terms with $n_a = 4n+1$ ($n_a = 4n+3$) are nonzero in class D (C). 
This is consistent with the relation in explicitly supersymmetric cSYK models, where for odd $N$ and supercharges that contain products of $n_a$ creation and annihilation operators one has $\mathcal{S} \mathcal{Q} \mathcal{S}^{-1} = (-1)^{(n_a-1)/2} \mathcal{Q}^\dagger$~\cite{Kanazawa:2017be}.
The fermion expansion structure is also similar to those of the two supercharges in classes D and C in the Majorana SYK model~\cite{Behrends2019}.

\section{Signatures in correlation functions}
\label{sec:signatures}

The zero modes and supersymmetry we find for odd $N$ have distinct long-time signatures in correlation functions.
Specifically, we show that, depending on the symmetry class and the fermion structure of the observable, a long-time plateau at nonzero (or conversely, strictly zero) value may emerge. 
A key to studying these plateaus is to consider observables connecting opposite charge sectors $q$ and~$-q$, i.e., to probe the inter-sector correlations. 
The plateaus we find are cSYK analogs of long-time plateaus discussed for the Majorana-SYK model~\cite{Cotler:2017fx,Behrends:2019jc,Behrends2019}.

We consider the retarded correlation function
\begin{equation}\label{eq:C_nq1}
 C_{nq}^+ (t) = -i \Theta (t) \frac{1}{N_{nq}} \sum_{\substack{ a,n_a=n, \\ q_a=q}} \langle \{ \bar{\Upsilon}_a (t) , \bar{\Upsilon}_a^\dagger (0) \} \rangle ,
\end{equation}
with the Heaviside step function $\Theta(t)$.
We average over all $N_{nq}$ different fermionic operators $\bar{\Upsilon}_a^\dagger$ that contain $n_a=n$ Majorana operators and increase the charge by $q_a=q$.
Although the operators $\bar{\Upsilon}_a$ are only defined up to a unitary transformation that keeps $n_a$ and $q_a$ intact, the correlation function defined above is invariant under such a transformation due to the sum over $a$.

\begin{figure}
 \includegraphics{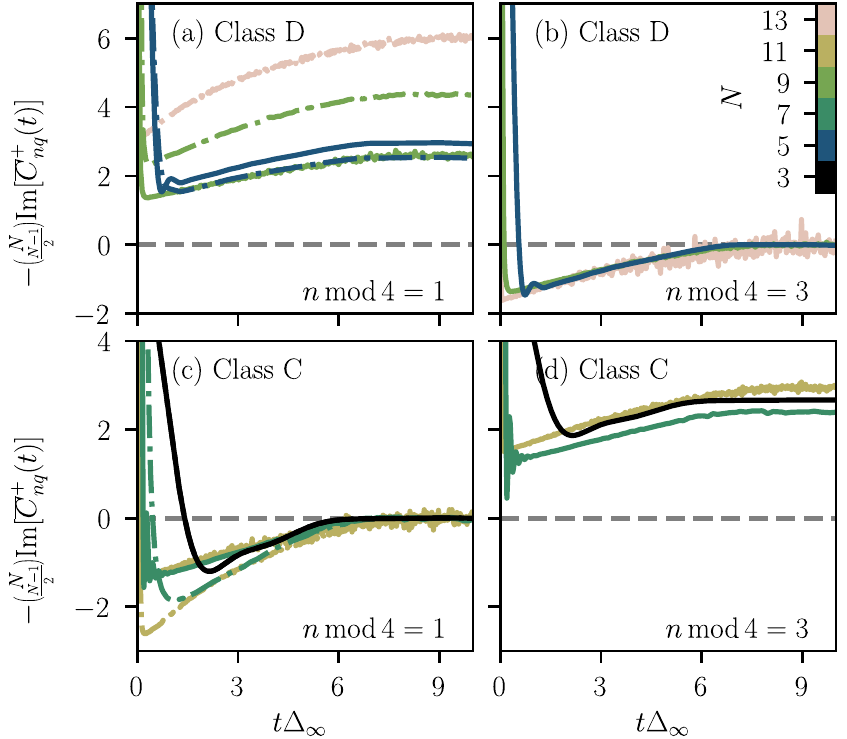}
 \caption{Infinite-temperature time-domain correlation function $\Imag [\overline{C}_{nq}^+(t)]$ with $q=1$ for various $N$ and number $n$ of creation/annihilation operators, averaged over an ensemble of up to $2^{16}$ normal-distributed $J_{ij,kl}$.
 The gray dashed lines in each panel show zero as a guide for the eyes. 
 The colors label the value of $N$ [cf.\ inset in panel~(b)].
 The $y$-axis is rescaled by the Hilbert space dimension of the charge sector for a more transparent comparison between different system sizes. 
 (a) Class D ($N \mod 4 = 1$) with $n \mod 4 = 1$, specifically $n=1$ (dashed-dotted lines) and $n=5$ (solid lines).  
 (b) Class D with $n = 3$.
 (c) Class C ($N \mod 4 = 3$) with ${n \mod 4 = 1}$; showing again $n=1$ (dashed-dotted lines) $n=5$ (solid lines).
 (d) Class C with $n =3 $. 
 }
 \label{fig:correlations}
\end{figure}

In Eq.~\eqref{eq:C_nq1}, angular brackets denote the quantum statistical average $\langle \ldots \rangle = \tr [ \ldots \rho_{q'} ]$ with respect to the thermal state in a given charge sector $\rho_{q'} = \mathcal{Z}_{q'}^{-1} \sum_\mu \exp (-\beta \varepsilon_\mu^{q'}) \ket{\psi_\mu^{q'}} \bra{\psi_\mu^{q'}}$. 
[The partition function is given by $\mathcal{Z}_{q'} = \sum_\mu \exp (-\beta \varepsilon_\mu^{q'})$.]
The value of $q'$ is linked to the charge $q$ in $C_{nq}^+ (t)$:
We want the action of $\bar{\Upsilon}_a^{(\dagger)}$ in Eq.~\eqref{eq:C_nq1} to include connecting opposite charge sectors; we must hence choose $q'=\pm q/2$. 
By inserting a resolution of the identity for the charge sectors $\bar{\Upsilon}_a^{(\dagger)}$ maps to, the correlation function can be rewritten as
\begin{align}
 C_{nq}^+ (t)
 =& - \frac{i \Theta(t)}{N_{nq} \mathcal{Z}_{q'}} \sum_{\substack{ \mu\nu,a\\ n_a=n\\q_a =q}} \left[
  \left| \bra{\psi_\mu^{q'}} \bar{\Upsilon}_a \ket{\psi_\nu^{q'+q}} \right|^2 e^{i t (\varepsilon_\mu^{q'}-\varepsilon_\nu^{q'+q})} \right. \nonumber \\
  & \left. 
+ \left| \bra{\psi_\mu^{q'}} \bar{\Upsilon}_a^\dagger \ket{\psi_\nu^{q'-q}} \right|^2 e^{ -i t (\varepsilon_\mu^{q'}-\varepsilon_\nu^{q'-q})} \right] e^{-\beta \varepsilon_\mu^{q'}}
\end{align}
where we used that $\bar{\Upsilon}_a^\dagger$ ($\bar{\Upsilon}_a$) increases (decreases) the charge by~$q$. 

\begin{figure}
 \includegraphics{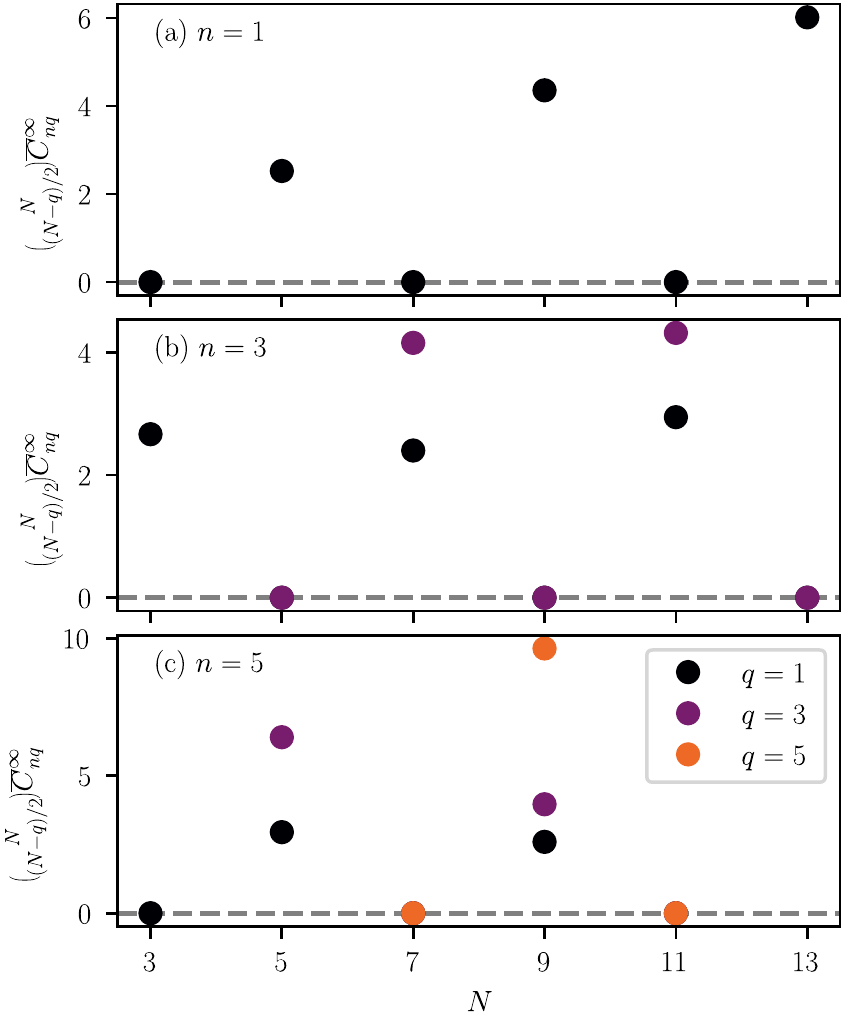}
 \caption{Values of the plateau $\overline{C}_{nq}^\infty$ as a function of $N$, averaged over an ensemble of up to $2^{15}$ normal-distributed $J_{ij,kl}$.
 In panels~(a) to~(c), we show the plateaus for different numbers $n$ of involved operators.
 The colors denote the change $q\le n$ in the charge.
 We rescaled all plateaus by the dimension of the corresponding charge sector for a better comparability between the different system sizes.
 }
 \label{fig:plateaus}
\end{figure}

Now separating terms with equal-energy opposite-charge matrix elements,  
we rewrite the correlation function as a sum of time-dependent and time-independent contributions,
\begin{align}
 C_{nq}^+ (t) &= - i \Theta(t) \left[ C_{nq}^\infty + \delta C_{nq}^+ (t) \right] \\
 C_{nq}^\infty &= \frac{1}{N_{nq} \mathcal{Z}_{q/2}} \sum_{\substack{ \mu,a\\ n_a=n\\q_a =q}}  e^{-\beta \varepsilon_\mu^{q/2}}
  \left| \bra{\psi_\mu^{-q/2}} \bar{\Upsilon}_a \ket{\psi_\mu^{q/2}} \right|^2 \nonumber .
\end{align}
The time-dependent contribution $\delta C_{nq}^+ (t)$ itself consists of two parts:
one contribution from charge sectors $q/2$ and $-q/2$ with degenerate energies $\varepsilon_\mu^{q/2} = \varepsilon_\mu^{-q/2}$ and hence \emph{cross-charge-sector} level repulsion, and another contribution from charge sectors $-q/2$ and $-3q/2$ (or $q/2$ and $3q/2$) that have uncorrelated energies.
Due to the lack of correlations, this second contribution quickly decays to zero. 
The first contribution, however, exhibits structure, including a ramp up to times $2\pi /\Delta_\infty$ (with the mean level spacing $\Delta_\infty$). 
This ramp is the result of the \emph{cross-charge-sector} spectral rigidity~\cite{Guhr:1998bg}, caused by the level repulsion.    
It connects to a plateau discussed below with a sharp corner, which is consistent with Gaussian unitary ensemble level statistics~\cite{Guhr:1998bg}.

We now discuss this long-time ($t \gg \Delta_\infty$) plateau. 
We focus on the infinite-temperature limit ($\beta \to 0$), where the origin and the features of the plateau are the most transparent. (The plateau is, however, present at any temperature, or even in a nonthermal correlation function with respect to an arbitrary eigenstate from $\rho_{\pm q/2}$.)
At $\beta \to 0$, the partition function simply counts the number of states in the charge sector, $\lim\limits_{\beta \to 0} \mathcal{Z}_{q/2} = \binom{N}{(N+q)/2}$, such that the value of the plateau can be rewritten as
\begin{align}
 C_{nq}^\infty &= \frac{1}{N_{nq} \binom{N}{(N+q)/2}} \sum_{\substack{ a, n_a=n\\q_a =q}} \sum_\mu \tr \left[ \bar{\Upsilon}_{a,q\mu}^\dagger \bar{\Upsilon}_{a,q\mu} \right]
\end{align}
with $\bar{\Upsilon}_{a,q\mu}  = P_{q\mu} \bar{\Upsilon}_a P_{q\mu}$ using the projector $P_{q\mu} = \ket{\psi_\mu^{q}} \bra{\psi_\mu^{q}} + \ket{\psi_\mu^{-q}} \bra{\psi_\mu^{-q}}$ on states with energy $\varepsilon_\mu^q = \varepsilon_\mu^{-q}$.
We can write $\bar{\Upsilon}_{a,q\mu}^\dagger$ in terms of the annihilation operator of the generalized fermion, $\bar{\Upsilon}_{a,q\mu}^\dagger = v_{q\mu}^a P_{q\mu} d^\dagger$ with the weight $v_{q\mu}^a = \bra{\psi_\mu^{q/2}} \bar{\Upsilon}_a^\dagger \ket{\psi_\mu^{-q/2}}$.
If $\bar{\Upsilon}_{a,q\mu}^\dagger \triangleq P_{q\mu} d^\dagger$, where $\triangleq$ means that the two sides transform in the same way under chiral symmetry , the weight $v_{q\mu}^a$ is generally nonzero, otherwise, $v_{q\mu}^a =0$.
Since $\mathcal{S}P_{q\mu}\mathcal{S}^{-1}=P_{q\mu}$, this implies that the value of the plateau
\begin{align}
 C_{nq}^\infty = 
 \begin{cases}
 \text{nonzero} & ~ \bar{\Upsilon}_{a}^\dagger \triangleq d^\dagger \\
 0 & ~ \text{otherwise.}
 \end{cases}
\end{align}
As noted for the fermion structure in Table~\ref{tab:transformation},  $\bar{\Upsilon}_{a}^\dagger \triangleq d^\dagger$ when $n_a \mod 4 = 1$ ($n_a \mod 4 = 3$) in class D (C).

To illustrate above findings, we compute ensemble-averaged correlation functions numerically.
To this end, we use the Hamiltonian $H_0 = (H + \mathcal{S} H \mathcal{S}^{-1})/2$ with $H$ from Eq.~\eqref{eq:hamiltonian}, and draw the complex couplings from a normal distribution that satisfies~\cite{Sachdev:2015dp,Davison:2017hf}
\begin{equation}
  \overline{ J_{ij,kl} J^*_{ij,kl} } = \frac{J^2}{8 N^3} ,
\end{equation}
where $\overline{(\ldots)}$ denotes the ensemble average.
The Hamiltonian $H_0$ satisfies chiral symmetry by construction; using it amounts to subtracting $\sum_{ijk} J_{ij,ki} \left( c_j^\dagger c_k - c_k c_j^\dagger \right)$ from Eq.~\eqref{eq:hamiltonian}.
In Fig.~\ref{fig:correlations}, we show $\Imag \overline{C}^+_{nq} (t)$ at infinite temperature. 
In all panels, we use the charge difference $q=1$, i.e., a density matrix in the charge sector $-q/2=-1/2$.
Fig.~\ref{fig:correlations}(a--b) show class D ($N \mod 4 = 1$). As argued above, for $n \mod 4 = 1$, the correlation function does not decay to zero, but rather approaches a nonzero plateau [panel (a)].
As the value of this plateau decreases exponentially with system size, we rescale the correlation function by the size of the charge sector.
When $n \mod 4 = 3$, however, the plateau value is zero [panel (b)].
Fig.~\ref{fig:correlations}(c--d) show class C ($N \mod 4 = 3$).
Here, the behavior is reversed compared to class D: The correlation function decays to zero when $n\mod 4 = 1$ [panel~(c)], but it approaches a nonzero plateau when $n\mod 4 = 3$ [panel~(d)].

In Fig.~\ref{fig:plateaus}, we show the ensemble averages $\overline{C}_{nq}^\infty$ for various system sizes $N$, charge sectors $q$, and numbers of creation/annihilation operators $n$, with $n=1$, $n=3$, and $n=5$ in panels (a), (b), and (c).
In all panels, we again rescale $\overline{C}_{nq}^\infty$ by the dimension of the charge sector $-q/2$.
Similarly to the correlation functions in the Majorana SYK model, the rescaled plateau for $n=1$ and $q=1$ increases with system size in the range we can simulate numerically~\cite{Behrends:2019jc,Behrends2019}.
We leave more detailed investigations of the plateau as a function of system size and $n$, $q$ for future work.

\section{Conclusion and Outlook}
\label{sec:conclusion}

In this work we have embedded the fourfold pattern of level spacing statistics~\cite{You:2017jj} for the chiral-symmetric complex-fermion SYK model into the Altland-Zirnbauer symmetry classification first introduced for its Majorana counterpart~\cite{Behrends:2019jc}.
While for an even number $N$ of fermions, this embedding results in the same classes (AI and AII) as previously recognized (as GOE and GSE) through studies of level spacing statistics, new classes (C and D), and consequences thereof, emerge for odd $N$. 
Below we first summarize these odd-$N$ consequences, both in terms of structural features and signatures. Then, we shall comment on hitherto unexplored features that may arise for even $N$ due to similar considerations.

We found that the odd-$N$, chiral-symmetric, cSYK model supports fermion-parity-odd many-body zero modes.
From these zero modes, we can build a generalized (many-body) fermion $d$ that does not contribute to the system's energy (i.e., as its constituent zero modes, commutes with the Hamiltonian).
Closely related to this generalized fermion, we find the emergence of $\mathcal{N}=2$ supersymmetry in the cSYK model.
This SUSY only relies on chiral symmetry, and, unlike previous explicitly supersymmetric cSYK constructions, requires no additional fine tuning of the couplings.

We linked the generalized fermion to signatures in correlation functions, in particular, long-time plateaus whose value is determined by the charge sector and symmetry class.
These signatures thus allow us to distinguish between classes C and D, which would appear the same through the lens of just level spacing statistics.
With potential condensed-matter~\cite{Chen:2018ho,Altland:2019ck} or quantum circuit realizations~\cite{Luo:2019fp,Babbush:2019eh} of the cSYK model in sight, these signatures may be useful for identifying and classifying the system in the mesoscopic regime.
In condensed matter realization of the SYK and cSYK models, the single-fermion ($n=1$) correlators (i.e., single-particle spectral function) can be measured using scanning tunneling spectroscopy~\cite{Pikulin:2017js}.
More complicated correlation functions may be measured in digital quantum simulations of the cSYK model~\cite{Luo:2019fp}.

The identification of the symmetry classes and their corresponding signatures in correlation functions is a strategy that may apply beyond the cSYK model.
Whenever different sectors of an interacting Hamiltonian are connected by an antiunitary symmetry, correlation functions of operators connecting these sectors may bear fingerprints of cross-sector correlations. 
In the cSYK model, this is especially transparent since the couplings between different fermions are structureless and no unitary symmetries besides charge conservation are present.
However, even for more structured models, e.g., perturbations around integrable points, cross-correlations between different sectors may be manifested in long-time limits of certain correlation functions.

Although we focused on $N$ odd, this strategy may lead to interesting new signatures in the even $N$ case as well. 
While in this case no generalized fermion that commutes with the Hamiltonian can be constructed and supersymmetry is not present,
charge sectors $q$ and $-q$ remain correlated in the presence of chiral symmetry.
Operators connecting these sectors must be thus be fermion-parity-even, which is the key difference from the odd $N$ case we focused on in this work.
Nevertheless, the correlation functions of these operators should still display signatures of cross-charge-sector level repulsion of the GUE type (cf.\ Table~\ref{tab:classes}) and of the existence of fermion-parity-even cross-charge-sector zero modes.

As discussed in Sec.~\ref{sec:model}, it is possible to use a more relaxed definition of chiral symmetry, in particular, to allow a transformation $\mathcal{S} H \mathcal{S}^{-1} = H + \varepsilon_0 Q$ with some energy $\varepsilon_0$.
In this case, the degeneracy of energy eigenvalues in opposite charge sectors is lifted.
The energies, however, remain related to each other by a constant that depends on the charge sector $\varepsilon_\mu^q - \varepsilon_\mu^{-q} = -q \varepsilon_0$.
In this case, SUSY is absent and the previously zero modes $\mathcal{O}^q_{\mu\mu}$ now  satisfy $[H,\mathcal{O}^q_{\mu\mu}]=-q \varepsilon_0 \mathcal{O}^q_{\mu\mu}$  [cf.\ Eq.~\eqref{eq:eigenops}].
However, the long-time plateau of the correlation function does not quite vanish, but rather changes to oscillations up to infinite times, with the amplitude given by the $\varepsilon_0=0$ plateau value and the frequency by $q \varepsilon_0$.

Another possible extension of the symmetry classification is the extension to $n$-body interactions.
Depending on the correlations in the couplings, chiral symmetry may still be present in a more generic $n$-body Hamiltonian.
Since the signatures we revealed in this work can be regarded as a direct consequence of chiral symmetry, we expect them to hold even for more generic Hamiltonians, although we leave the careful investigation of this for future studies. 

\begin{acknowledgments}
We thank A.~Chen, M.~Franz and D.~Pikulin for helpful discussions.
This work was supported by the ERC Starting Grant No. 678795 TopInSy.
\end{acknowledgments}

\bibliography{complex_sy}

\end{document}